\documentclass{mem}
\usepackage{natbib}\usepackage{txfonts}\usepackage{balance}
\usepackage{graphicx}
\usepackage[a4paper]{hyperref}
\usepackage{epstopdf}
\begin{document}
\def\teff{$T\rm_{eff }$}
\def\kms{$\mathrm {km s}^{-1}$}

\title{
Cosmological Numerical Simulations of Radio Relics in Galaxy Clusters: Insights for Future Observations
}

   \subtitle{}

\author{
Jack O. \,Burns\inst{1,2} 
\and Samuel W. \,Skillman\inst{1,3}
          }


\institute{
Center for Astrophysics \& Space Astronomy,
University of Colorado at Boulder,
Boulder, CO 80309 USA
\and
NASA Lunar Science Institute, Ames Research Center,
Moffett Field, CA 94035 USA
\and
DOE Computational Science Graduate Fellow \\
\email{jack.burns@cu.edu}
}

\authorrunning{Burns \& Skillman}

\titlerunning{Simulations of Radio Relics}

\abstract{The acceleration of electrons at shock fronts is thought to be responsible for radio relics, extended radio features in the vicinity of merging galaxy clusters.  By combining high resolution Adaptive Mesh Refinement Hydro/N-body cosmological simulations with an accurate shock-finding algorithm and a model for electron acceleration, we calculate the expected synchrotron emission resulting from cosmological structure formation.  From these simulations, we produce radio, SZE and X-ray images for a large sample of galaxy clusters along with radio luminosity functions and scaling relationships.  We find that with upcoming radio arrays, we expect to see an abundance of radio emission associated with merger shocks in the intracluster medium.  By producing observationally motivated statistics, we provide predictions that can be compared with observations to further our understanding of electron shock acceleration and kinematic structure of galaxy clusters.

}
\maketitle{}

\section{Introduction}
Colliding galaxy clusters are insightful astrophysical plasma laboratories.  Shocks produced during mergers heat the intracluster medium (ICM), assisting the gas to achieve roughly hydrostatic equilibrium with the cluster gravitational potential well.  Shocks also play a key role for the nonthermal component of the ICM.  Shocks compress and amplify magnetic fields.  Shocks accelerate cosmic rays (CR) via a diffusive Fermi process.  Thus, merger shocks are illuminated via the resulting synchrotron radiation arising from relativistic CR electrons gyrating in ICM B-fields producing steep-spectrum, so-called ``radio relics''.

\begin{figure*}[t!]
\centering
\resizebox{10.0cm}{!}{\includegraphics[clip=true]{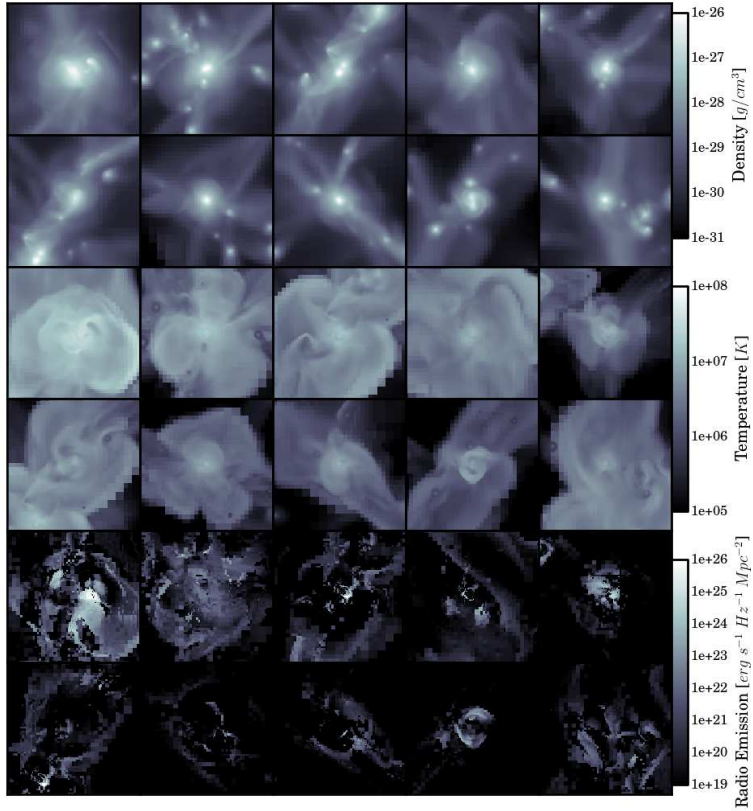}}

\caption{\footnotesize
Projections of density (top), temperature (middle), and 1.4 GHz radio emission (bottom) for a representative sample of clusters from {\it Enzo} AMR simulations at $z=0$.  Each image is 4 $h^{-1}$ Mpc on a side.  Peak resolution is 3.9 $h^{-1}$ kpc.
}
\label{3panel}
\end{figure*}

Radio relics are rarely found in radio surveys of galaxy clusters, possibly because of their steep radio spectrum and diffuse, low surface brightness emission at cm wavelengths.  However, with new and improved radio arrays such as LOFAR, GMRT, and the EVLA, new high sensitivity observations may reveal an abundance of cluster radio relics.  To assist in searching for these relics, we have run high spatial dynamic range cosmological adaptive-mesh-refinement (AMR) simulations using the {\it Enzo} code that produce $\approx$2000 clusters in various stages of merger evolution \citep{skillman10}.  Using new shock-identification tools and analytical models for diffusive shock acceleration \citep{hoeft07}, we have constructed radio maps of a large sample of simulated clusters to study the production and observational properties of relics in different merger states.  Details of these {\it Enzo} simulations are found in \cite{skillman10}.

\section{Simulated Radio Relics}

Images of density, temperature, and radio emission for a small representative sample of clusters in a single projection along one axis are shown in Figure \ref{3panel}.  Several interesting points are worth noting in comparing these images.  First, there is a distinct difference in morphologies between the density and radio images.  The density ($\approx$X-ray emission) is center-filled whereas the radio is often edge-brightened.  The curved radio arcs are illuminated bow shocks produced as two clusters pass between their cores.  Such shocks are effectively invisible on the density/X-ray maps and only partially visible on the temperature images.  Thus, the radio relics light up important evolutionary features in clusters (i.e., shocks) that are not apparent at X-ray energies.

Second, in cases where the merger is largely along the plane of the sky, the radio relics fall on the edges of sharp temperature gradients.  This is particularly apparent for the cluster in the lower right of Figure \ref{3panel}.  This strong X-ray temperature/radio spatial correlation agrees with recent observations.

\begin{figure}[]
\resizebox{\hsize}{!}{\includegraphics[clip=true]{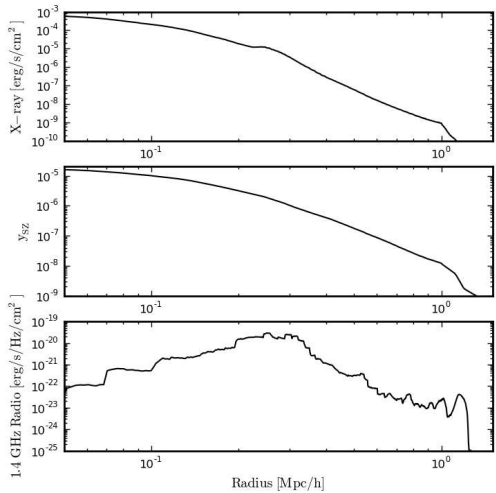}}
\caption{\footnotesize Profiles of X-ray flux, dimensionless Compton y SZE parameter ($\propto$ gas pressure), and radio flux density for a merging cluster with a center-filled radio relic in the upper left-hand image of the lower panel in Figure \ref{3panel}.
}
\label{Figure2}
\end{figure}

Third, depending upon the projection, there is a wide variety of radio morphologies.  Although there is a preference for edge-brightened radio emission coinciding with shocks, there are a few clusters that demonstrate more diffuse, center-filled radio emission (upper left-hand clusters in bottom panel of Figure \ref{3panel}).  This corresponds to cluster projections where the merger is largely along the line-of-sight.  Such emission may qualitatively resemble ``radio halos'', but they do not demonstrate a strong correlation between X-ray and radio emission profiles seen for radio halos.  In Figure \ref{Figure2}, we show profiles of X-ray, Sunyaev-Zeldovich Effect (SZE), and radio emission for a cluster with apparently center-filled radio relic emission. For all the radio relics, the radio profile shapes and slopes are inconsistent with those for the densities (and X-ray emission). So, one should be able to distinguish between real radio halos and projected relics via their X-ray/radio profiles, along with expected differences in spectral index.

\section{Scaling Relations}

\begin{figure}[]
\resizebox{\hsize}{!}{\includegraphics[clip=true]{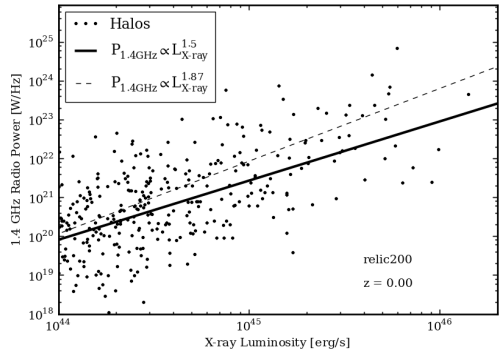}}
\caption{\footnotesize Scaling relation between 1.4 GHz radio power and $0.2-12$ keV X-ray luminosity for simulated clusters (solid line).  Also shown (dashed line) is a line with the slope of the best-fit observed scaling relation for radio relics from \cite{feretti02}.
}
\label{Figure3}
\end{figure}

In Fig. \ref{Figure3}, we show the scaling relation between radio power at 1.4 GHz and soft X-ray luminosity for the most X-ray luminous simulated clusters in our sample.  Although there is a clear correlation between the radio and X-ray luminosities within $r_{200} \approx 0.8r_{virial}$, there is considerable scatter in the relationship (factor of $\approx 10^4$ in $P_{1.4 GHz}$).  This scatter is real and represents different merger states for different clusters.  High radio power clusters have suffered recent mergers whereas low power clusters have not experienced a merger in over a Gyr.  To date, only the most radio luminous clusters have been observed and this possibly represents an observational bias that may be remedied with more sensitive, high bandwidth observations in the near future.

There is also good agreement in the slopes of the scaling relation between observed and simulated clusters in Fig. \ref{Figure3}. Although the observed sample from \cite{feretti02} is small (9 clusters), the general agreement between simulations and observations is encouraging.

\section{Predicted Number of Radio Relics}

From our numerical simulations, we constructed a radio luminosity function for radio relic clusters.  That is, we calculated the cumulative number of clusters with $P_{1.4 GHz}$ greater than a given level, where the radio emission is the result of cluster merger shocks.  Since radio power scales with cluster mass in our simulations ($P_{1.4 GHz} \propto M^{3.2}$) \citep{skillman10}, our computational volume produces clusters with $M_{200} < 10^{15}$M$_\odot$ and $P_{1.4 GHz} < 10^{24}$ W/Hz.  To extend these limits to higher masses and radio powers, we extrapolated our cluster sample using the \cite{warren06} mass function at $0<z<1$. 
Within this cosmological volume of $\approx$26 (Gpc/h)$^3$, our simulations and extrapolations predict that $\approx$200 to 1000 radio relic clusters with integrated (over an area with $r=r_{200}$) $ P_{1.4 GHz}>10^{25}$ W/Hz will be present.  With a factor of 10 improvement in the sensitivity of the EVLA at 1.4 GHz, we expect to see 10-100 times more galaxy clusters with radio relics.  For an all-sky survey out to $z \approx 0.5$, there should be $\approx 200$ clusters with $P_{1.4 GHz} > 10^{25}$ W/Hz. 

\section{Shocks on SZE Images}

\begin{figure}[]
\resizebox{\hsize}{!}{\includegraphics[clip=true]{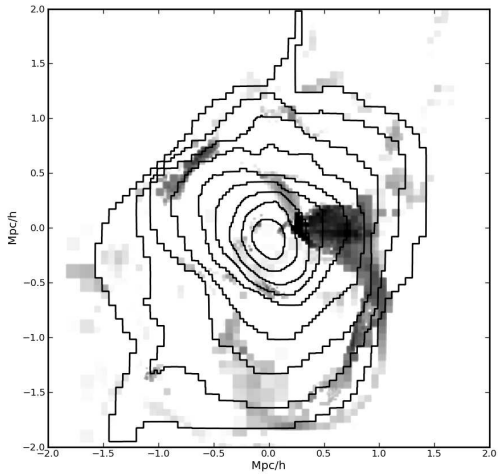}}
\caption{\footnotesize Synthetic SZE/radio images.  Contours are SZE flux and grey-scale is radio emission.
}
\label{Figure4}
\end{figure}

Fig. \ref{Figure4} shows an overlay of radio onto SZE images for one cluster in our simulated sample of clusters.  Note that the radio relic shocks correlate with sharp gradients on the SZE map.  Also, there is extended SZE structure corresponding to the strongest relic emission.  This may suggest that as observations continue to improve, SZE will complement radio images in mapping merger shocks.

\section{Conclusions}

Using a robust shock-finding algorithm and an analytical model for diffusive shock acceleration applied to a large-volume AMR cosmological simulation, we have produced synthetic X-ray and radio images with characteristics similar to those observed in clusters with radio relics.  The dual-arc radio morphologies and X-ray/radio scaling relations are good matches to observations.  We predict that an increase in 10-100 in the number of radio relic clusters detected with new and improved radio arrays (e.g., EVLA, LOFAR, and future lunar farside low frequency telescopes) should be possible.

\begin{acknowledgements}
This work has been funded by grants from the U.S. NSF (AST-0807215) and the NASA Lunar Science Institute (NNA09DB30A) to J.O.B.  S.W.S. has been supported by a DOE Computational Science Graduate Fellowship (DE-FG02-97ER25308).  Computations described in this work were performed using the Enzo code developed by the Laboratory for Computational Astrophysics at the University of California in San Diego (http://lca.ucsd.edu).  Analysis of the simulations was performed using \textit{yt} \citep{turk11}. 
\end{acknowledgements}

\bibliographystyle{aa}

\end{document}